\documentclass[conference]{IEEEtran}
\usepackage{cite}
\usepackage{amsmath,amssymb,amsfonts, bbm}
\usepackage{algorithmic}
\usepackage{graphicx}
\usepackage{xcolor}

\def\BibTeX{{\rm B\kern-.05em{\sc i\kern-.025em b}\kern-.08em
    T\kern-.1667em\lower.7ex\hbox{E}\kern-.125emX}}
\def\BibTeX{{\rm B\kern-.05em{\sc i\kern-.025em b}\kern-.08em T\kern-.1667em\lower.7ex\hbox{E}\kern-.125emX}}

\DeclareMathOperator*{\clip}{clip}
\title{DeepADMR: A Deep Learning based Anomaly Detection for MANET Routing}
\author{Alex Yahja \qquad 
  Saeed Kaviani \qquad
  Bo Ryu \qquad \quad 
  Jae Kim \qquad \quad
  Kevin Larson \\
  Episys Science Inc, Poway, CA \qquad and \qquad The Boeing Company, Seattle, WA  \\
  \small{\texttt{\{alex, saeed, boryu\}@episci.com} \texttt{\{jae.h.kim, kevin.a.larson\}@boeing.com}} }
\begin{document}
\maketitle
\thispagestyle{plain}
\pagestyle{plain}
\begin{abstract}
We developed DeepADMR, a novel neural anomaly detector for the deep reinforcement learning (DRL)-based DeepCQ+ MANET routing policy. The performance of DRL-based algorithms such as DeepCQ+ is only verified within the trained and tested environments, hence their deployment in the tactical domain induces high risks. DeepADMR monitors unexpected behavior of the DeepCQ+ policy based on the temporal difference errors (TD-errors) in real-time and detects anomaly scenarios with empirical and non-parametric cumulative-sum statistics. The DeepCQ+ design via multi-agent weight-sharing proximal policy optimization (PPO) is slightly modified to enable the real-time estimation of the TD-errors. We report the DeepADMR performance in the presence of channel disruptions, high mobility levels, and network sizes beyond the training environments, which shows its effectiveness. 


\end{abstract}
\section{Introduction}
Deep reinforcement learning (DRL) algorithms are successfully applied to complex high-dimensional problems, mainly due to the use of deep neural networks (DNN) for function approximations \cite{mnih2015human}. Researchers have applied DRL algorithms for various problems in mobile ad-hoc networks (MANETs), e.g., for minimizing average or worst-case end-to-end delay in routing problems \cite{choi2021deep, keum2022trust} and routing path optimization \cite{dong2018intelligent}. In \cite{kaviani2021deepcq+}, it is shown that the DRL-based DeepCQ+ algorithm outperforms the state-of-the-art robust routing for dynamic networks (R2DN) \cite{johnston18}. In this context, robust routing relies on the routing of the unicast data flows with the availability of simplified multicast forwarding (SMF) (as simplified data flooding and relaying) among routing peers \cite{macker2005simplified}. 

Despite the generalization ability of learned DRL-based policy, however, applying these algorithms to different environments necessitates re-learning to attain similar performance, with methods such as curriculum learning \cite{bengio2009curriculum}, data augmentation \cite{raileanu2021automatic}, and policy distillation \cite{rusu2016policy}. DRL agents endeavor to generalize in perturbed and adversarial environments, incurring high risk in the deployment of the trained policy in the tactical domain where reliability and human-operator trust is vital. Hence, a reliable anomaly detection method is vital. Indeed, any deployment of DRL agents without characterizing their measurement errors and thus detecting and characterizing anomalies, as the physical world and physics require, can incur mission failures.

Eschewing centralized command and control, MANETs present challenging network routing problems in dynamically changing environments. MANET nodes can die and their links can attenuate, and their domains are prone to shift to new regimes. These incur inaccuracies and breakdowns in the performance of DRL algorithms, which is known (about the unknown distributions outside training data distributions) for any learning and adaptive system, humans included. The adage that ``no plan survives the first contact with the enemy'' cautions against any confidence in the MANET schemes and DRL training results sans expert-constructed diagnostics and sanity enforcement. MANETs are often canvased with anomalies that matter impacting mission success. 

Anomaly refers to deviation from the expected operational norm. An anomaly can be irregular and rare and typically have unknown distribution. Anomaly detection is typically performed by measuring how much the data points lie outside the nominal range in a persistent manner, that is, persistent outliers. Due to the nature of RF propagation (especially for mmWave of 5G), user device movements/platform mobility, and environmental disturbance, MANET failures are frequent and this makes anomaly detection the core enabler for continuous MANET operations. This implies that the performance metric for MANETs should be the resiliency score, which is computed by anomaly detectors. The resiliency score for MANETs dominates in importance over other metrics such as overhead/efficiency and goodput carried over from fixed network metrics. Indeed, frequent MANET failures necessitate anomaly detection functionality in dynamic network operations in tactical environments.

This paper describes our novel neural anomaly detection method. We leverage the advances in Multi-Agent DRL to imbue learned agent policies into human-crafted rules and into anomaly detection. Specifically, our anomaly detection method, DeepADMR, is based on temporal difference error (TD-Error) \cite{sutton2018reinforcement} and cumulative sum (CUSUM)-like \cite{kurt2019sequential} methods. We applied the anomaly detector to our DeepCQ+ \cite{kaviani2021deepcq+} routing testbed. We chose TD-Error as our metric for resiliency as it is fundamental and effective \cite{van2016true,sutton2018reinforcement}, and the distributional TD-Error is shown to be associated with dopamine-based learning in the brain \cite{dabney2020distributional}. Our experiment results on anomalous network conditions such as channel disruptions, network size changes, and node mobility changes show the effectiveness of our approach.

In many domains, physics- and expert-elicited rule-based algorithms exist. DRL agents can complement but not replace these expert-crafted models or rules; in our case, reliable network routing management. Indeed, DRL has the unsolved 1\% problem, a dangerous void such that it cannot ensure safe and effective operations in new, adversarial, and perturbed environments, and hence our DeepADMR anomaly detection and diagnostics fills this primary need. 

The contributions of this work are as follows: 
\begin{enumerate}
\item We summarize our DeepCQ+ neural routing \cite{kaviani2021deepcq+} with modification and training regime for anomaly detection.
\item We describe our DeepADMR, a new real-time non-parametric neural anomaly detection method based on empirical cumulative-sum-like monitoring of TD-Error streams to produce a learned policy for novel, adversarial and perturbed environments.
\item We present the results showing the effectiveness of DeepADMR in real-time anomaly detection of network behavior.
\end{enumerate}

\section{Related Work}
Inverse of Wasserstein generative adversarial network (WGAN)  \cite{arjovsky2017wasserstein} is used to detect anomaly in which an original, nominal distribution is mapped into the normal distribution in the training phase \cite{mestav2020universal}. The resulting normal distribution is discretized into uniform cell symbols (``buckets'' or discrete partitions) according to the Birthday Coincidence uniformity test for universal data anomaly detection\cite{mestav2020universal}. In the execution phase, a new distribution is fed into the learned Inverse WGAN for the coincidence uniformity test to decide on whether the new distribution is normal or anomalous based on certain threshold. 

Actor-Critic neural network is used to detect network anomaly with nominal data point of sensor data series is given the reward of 0 and the abnormal sensor data point is instead given the reward of 1. Formulating this sequence of sensor readings as Markov Decision Process, the confidence level that a hypothesis is true is then computed as Bayesian log-likelihood ratio \cite{joseph2021scalable} of Actor-Critic policy.

\section{Temporal Difference Error}
In DRL, the target environment is often modeled as a Markov Decision Process (MDP). We consider a MDP $(\mathcal{S};\mathcal A;\mathcal P;\mathcal R; \gamma)$ for some given state space ($\mathcal S$), action space ($\mathcal {A}$), transition dynamics ($\mathcal T$), reward function ($\mathcal R$) and discount factor ($\gamma$). For a given and deterministic policy $\pi: \mathcal{S} \mapsto \mathcal A$, the action value function $Q^\pi$ at time-step $t$ is defined as the expected cumulative reward under the policy starting from state $s_t$ with action $a_t$ during a time horizon of $T$ as
\begin{equation}\label{action-value}
\begin{split}
    Q^\pi(s_t,a_t) &:= \mathbb{E}_t^\pi\left[\sum_{l=0}^T \gamma^l r_{t+l} \bigg | s_0 = s, a_0 = a\right] \\ &=  \mathbb{E}_t^{\pi}[r_t+\gamma Q^\pi(s_{t+1},\pi(s_{t+1}))],
\end{split}
\end{equation}
where the expectation $\mathbb{E}_t^\pi$ is over samples of $r_t \sim \mathcal R(s_t,a_t)$ and $s_{t+1} \sim \mathcal P(s_t,a_t)$. $\gamma$ is the discount factor for the cumulative reward. The action-value function can be recursively calculated as (\ref{action-value}) where it describes the well-known Bellman equation \cite{sutton1988learning}. 

\begin{figure}[ht]
\centering
\includegraphics[width=0.45\textwidth]{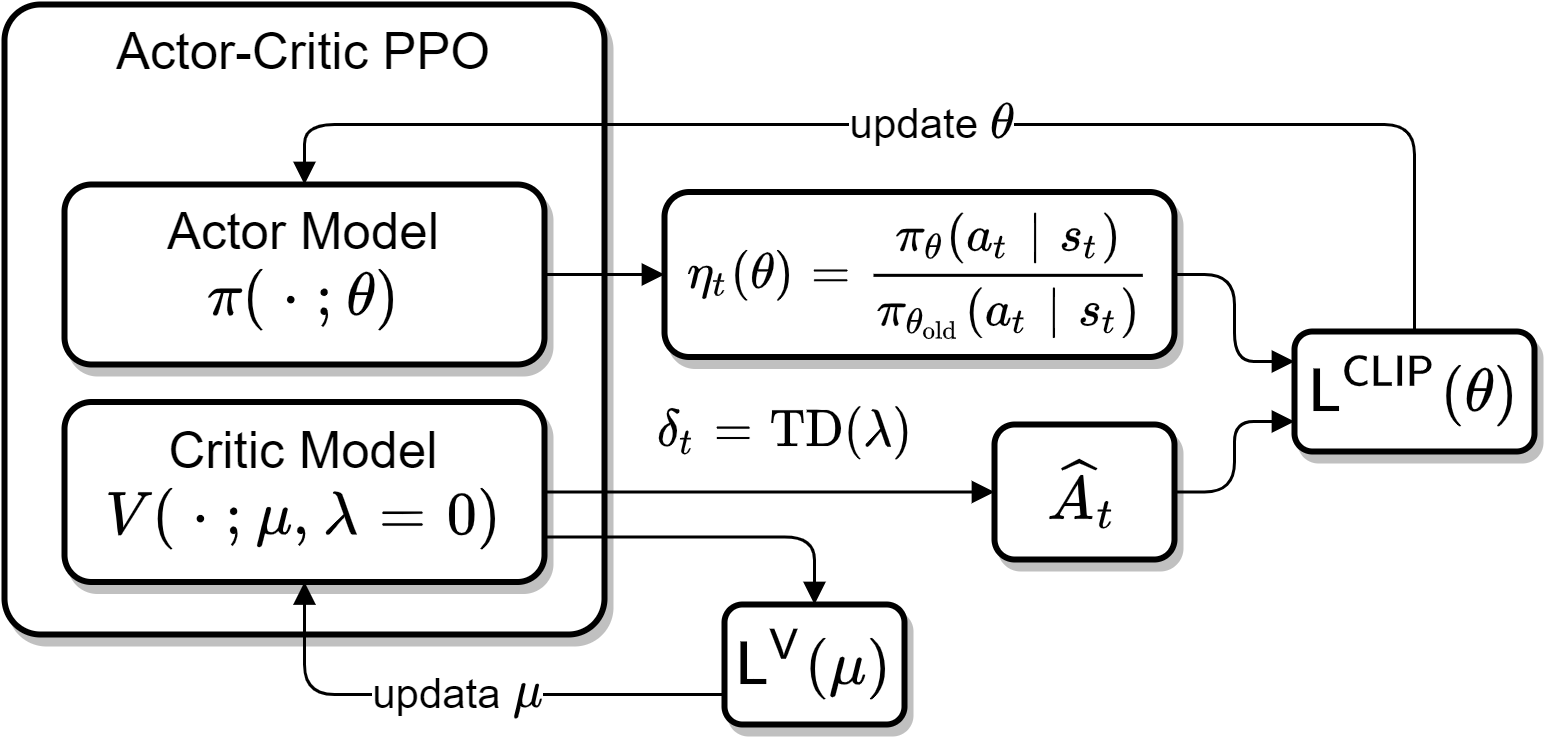}
\caption{\label{fig:ppo} Actor and critic models trained separately in PPO training. The critic model is trained using GAE parameter $\lambda = 0$ for the use in DeepADMR.}
\end{figure}

Our DRL-based DeepCQ+ routing policy \cite{kaviani2021deepcq+} deploys a weight-sharing proximal policy optimization (PPO) \cite{schulman2017proximal} as our learning algorithm as it trains robustly over parameter space. PPO uses the expected return at each node's state (i.e. value estimate $V(s_t) = \max_{a}Q(s_t, a)$). We minimize Temporal Difference error, TD-Error $\delta_t$, the difference between value estimate at time $t$ (i.e. $V(s_t)$) and a possibly-better value estimate at the next time step $t+1$ (i.e. $r_t + \gamma V(s_{t+1}$) \cite{sutton2018reinforcement} and given by
\begin{equation}
    \delta_t = r_{t} + \gamma V(s_{t+1}) - V(s_t).
\end{equation}
 The objective function of PPO is as follows, 
\begin{equation}
\mathsf{L^{CLIP}}(\theta) = \mathbb{E}_t\!\left[\min\left(\eta_{\theta_t}\hat{A}_t, \clip\!\left(\eta_{\theta_t},1-\epsilon, 1+\epsilon\right)\!\hat{A}_t\right)\right] 
\end{equation}
where $\eta_\theta = \frac{\pi_\theta (a_t|s_t)}{\pi_{\theta_\text{old}}(a_t|s_t)}$ and $\theta_{\text{old}}$ is the vector of policy parameters before the update. $\hat{A}_t$ is the Advantage function using generalized advantage estimate (GAE) with $\lambda$ parameter given as,   
\begin{equation}
    \hat{A}_t = \delta_t + (\gamma\lambda)\delta_{t+1} + \ldots + (\gamma \lambda)^{T-t+1}\delta_{T-1}
\end{equation}
This estimator equals the $\text{TD}(\lambda)$ error estimator. When $\lambda = 0$, the estimator is equivalent to the TD-Error (i.e. $\text{TD}(0) = \delta_t$). Fig. \ref{fig:ppo} shows the PPO training,  with the two actor and critic models. The critic model, which is jointly trained using $\mathsf{L^V}(\mu)$ over parameters $\mu$, estimates the value function and is used for the computation of the TD-Errors. 

During execution, this TD-Error can be computed in real-time using the value network. The value network is the DNN that is trained using PPO to estimate the value function. Our training expectation is that the TD-Error to be small enough within the class of nominal training environments. If this value is not small, then it suggests that either the next node's state is different from expected and/or the reward signal from the environment is different from those expected during training. Intuitively, we expect TD-Error to be an indicator of irregular behavior, or surprise, of new environments and anomaly scenarios when monitored during execution, when observation mismatches expectation. 


\section{Our MANET Learning Testbed}
DeepCQ+ routing is described in \cite{kaviani2021deepcq+,kaviani2021robust}, where we make a DNN version for packet routing policy in MANETs. The anomaly detection in this work complements the DeepCQ+ routing; so it is crucial to describe the DeepCQ+ framework as the host platform for our anomaly detection.

DRL agents select an action based on a policy and observe the network response emitted by a reward. The reward acts as a reinforcement signal to improve the policy. Reinforcement learning (RL) aims to maximize the rewards over time, leading to optimal policy. In this case, the MANET nodes are considered to have SMF protocol and our routing optimization is on top of SMF protocol. In this context, agents (nodes) decide to which node(s) to forward packets, and choose to broadcast, or unicast (and to which next-hop node). The actions at each node (unicast or broadcast) are made by the routing policy given the state of the node. Similar to \cite{johnston18, kumar98} nodes keep a table of RL-metrics for node status described by two RL-metrics: $C$-factors and $Q$-factors. Each node $i$ keeps a $Q$-factor, $q(i,j,d)$ for each destination $d$ and potential next-hop $j$, which is an estimate of the quality of path to reach $d$ through neighboring node $j$ \cite{Boyan1994}. The $C$-factors are defined to oversee the dynamic of the network \cite{kumar98}. Indeed, $c(i,j,d)$ is a confidence metric representing the likelihood of a node $i$ reaching the destination $d$ eventually by using the next hop $j$. A detailed treatment of these two factors is given in our DeepCQ+ papers \cite{kaviani2021deepcq+,kaviani2021robust}. In DeepCQ+, $CQ$-vectors is the main component of the agent's state feed into the routing policy. 




\begin{figure}[ht]
\centering
\includegraphics[width=0.485\textwidth]{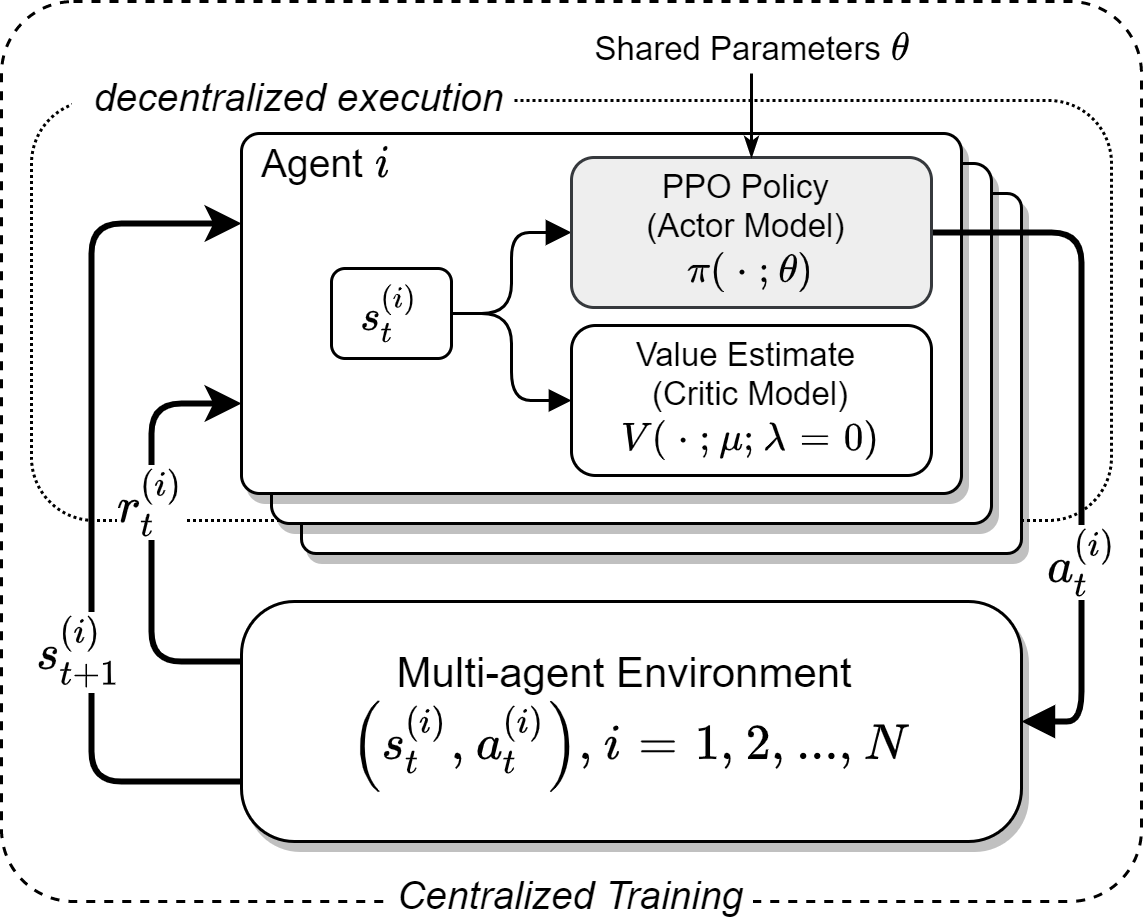}
\caption{\label{fig:ctde} Multi-agent network routing learner with shared policy parameters between agents with centralized training and decentralized execution. Each agent $i$, uses the shared policy $\pi_\theta$ individually to find its own action $a_i(t)$ based on its own observations $o_i(t)$. The multi-agent environment transitions to next state $s_{t+1}$ based on a combination of individual actions and emits rewards accordingly.}
\end{figure}

\begin{figure}[ht]
\centering
\includegraphics[width=0.485\textwidth]{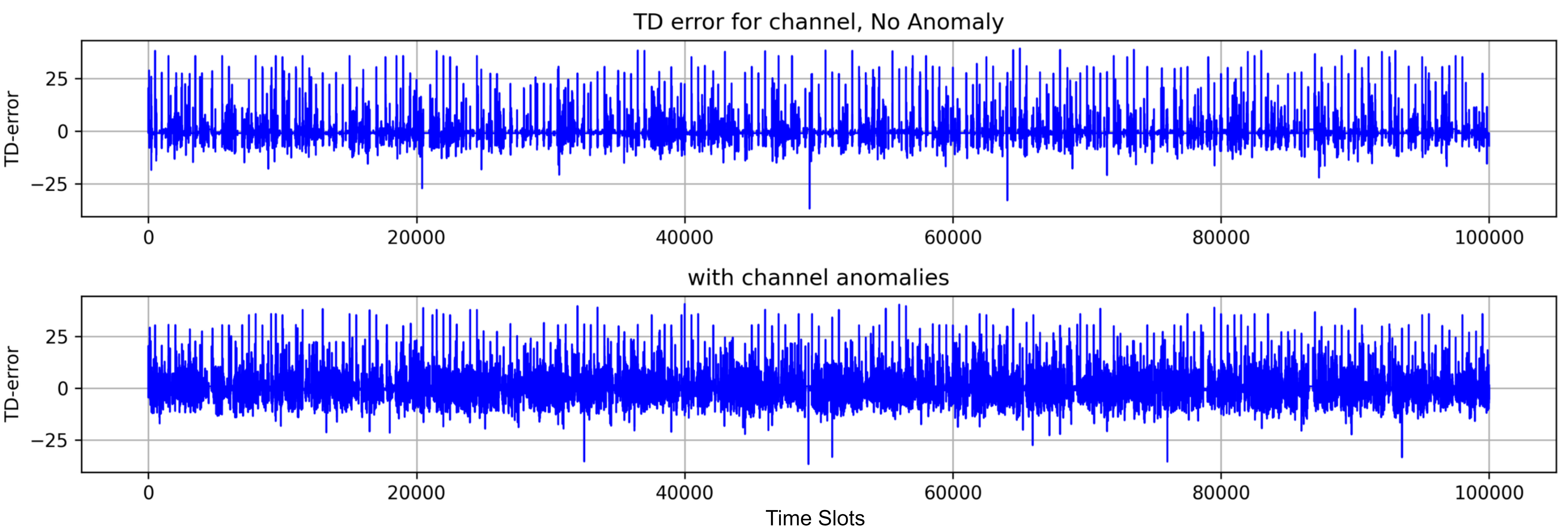}
\caption{\label{fig:td_error_ch_ad}Persistent change in TD-Error in bottom figure when MANET channel is disturbed by jammers compared to the anomaly-free operation on top.}
\end{figure}

We consider a homogeneous MANET with network size, $N$. This network holds multiple unicast traffic flows with randomized source and destination pairs. This means that incoming packets are injected at random nodes intended for random destinations and at random rates. The network dynamics are modeled as a movement of the nodes at various random velocities and directions. We use the Gauss-Markov as MANET mobility model. The Gauss-Markov mobility model covers the impact of network dynamics \cite{ariyakhajorn2006comparative}. Each realization of a network scenario (also called an \textit{episode} in the context of RL) has at least $T$ time periods (each time period is a single packet duration time). For simplicity, we assume that packet duration is fixed in time (i.e., a slotted system) but data rates can still vary. An episode ends when it runs over a maximum traffic length, $T_{max}$. 
The nodes have a duplicate packet detection (DPD) process to prevent routing packets from loops. 

We use PPO \cite{schulman2017proximal} for neural network training, with multiple epochs of stochastic gradient descent to update routing policy. Partial observability and communication constraints in MANET necessitate the learning of decentralized policies, which rely only on local information at each agent. Decentralized policies also obviate the exponential joint-action space, endowing them with fast convergence in training. Amazingly, decentralized policies can be learned in a centralized fashion via the paradigm of \textit{centralized training with decentralized execution}. Homogeneous agents, moreover, can share their policy network weights during training \cite{terry2020parameter}. Fig. \ref{fig:ctde} shows our multi-agent learner with centralized training, decentralized execution, and parameter sharing among agents. 

\section{Novel Neural Anomaly Detection}
Outlier detectors are sensitive to false alarms since it is possible to observe non-persistent random outliers under normal system operations. If a system exhibits persistent outliers, however, then this may indicate an actual anomaly. Hence, with anomaly defined as persistent outliers, we accumulate statistical evidence for anomaly over time, similar (but distinct in our non-parametric requisition) to the accumulation of log-likelihood ratios (LLRs) in the well-known CUSUM algorithm \cite{poor2008quickest} for change detection. These persistent outliers or changes are shown in Fig. \ref{fig:td_error_ch_ad}, where TD-Error values are compared non-parametrically for anomaly-free vs channel-disrupted operations in MANET.

\begin{figure}[ht]
\centering
\includegraphics[width=0.48\textwidth]{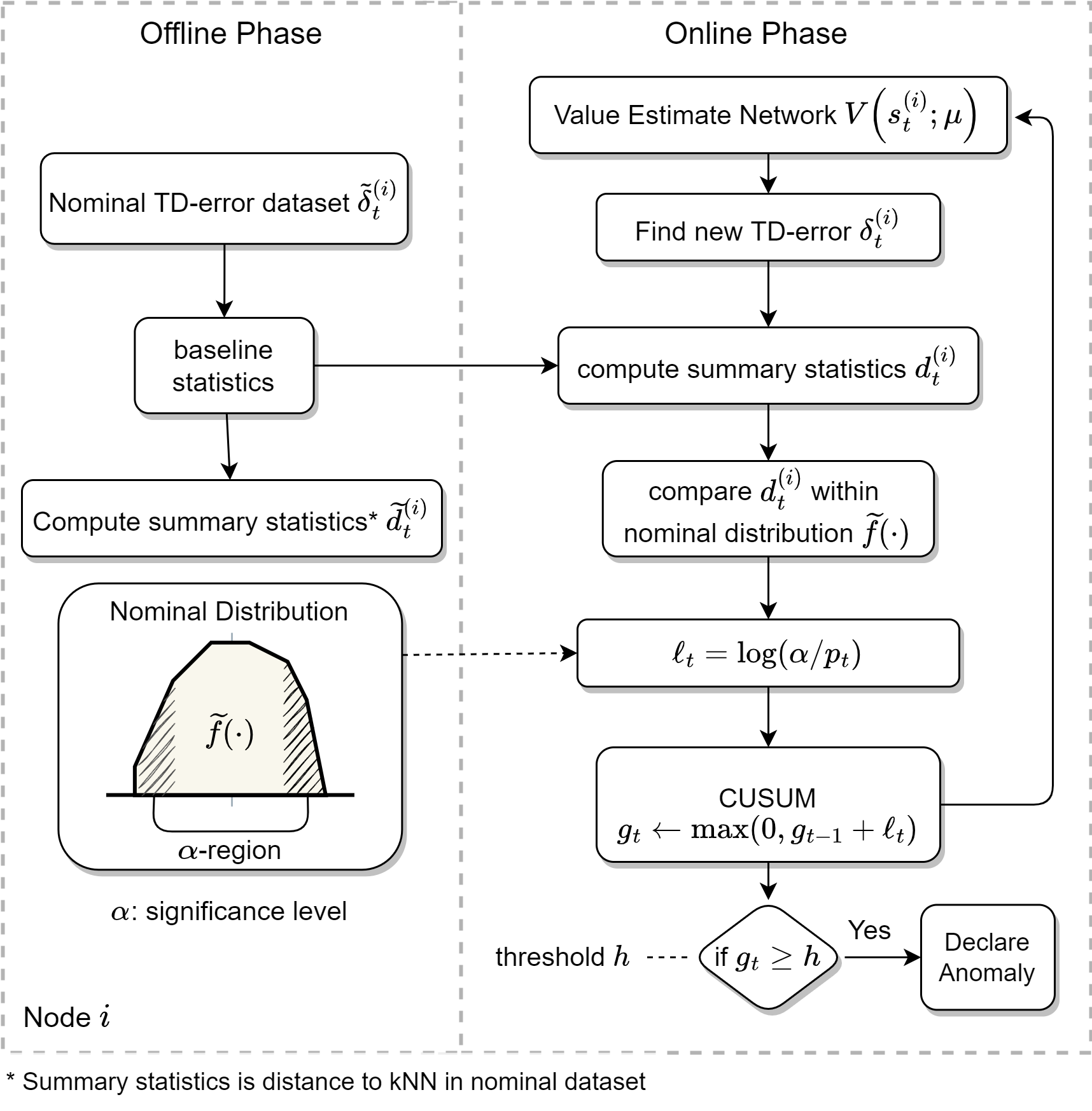}
\caption{\label{fig:offline_online}Offline and online phases of the anomaly detection.}
\end{figure}

As described, our novel real-time anomaly detection, DeepADMR, is based on the method of measuring temporal difference error, TD-Error \cite{sutton2018reinforcement}, and neural network learning. DeepADMR is diagrammed in Fig. \ref{fig:offline_online} and works as follows: 

First, we train our PPO learning agents with shared weights independently. We use locally available rewards for the PPO policy training to estimate the deviation from expected behavior locally and in a decentralized fashion. For this PPO policy training, we set GAE parameter $\lambda = 0$. The reason we train PPO with GAE parameter $\lambda = 0$ is that GAE advantage estimator is equivalent to the $\text{TD}(\lambda)$ error estimator. Using the $\text{TD}(\lambda)$ error to regularize actor-critic methods is equivalent to regularization with the GAE estimator with $\lambda$. Hence, when $\lambda = 0$, it becomes $\text{TD}(0)$ and this allows us to make use of TD-Error as the metric whenever PPO is trained with $\lambda = 0$ in its GAE. The advantage of GAE makes it less prone to overfitting environmental instance-specific features and thus leads to better generalization. In addition to learning the policy network, we also learn the value estimate network which estimates the value function $V(s_t; \mu)$ via trained weights $\mu$ (which is different from policy network weights $\theta$). Since the reward is considered available locally, we use this value network to estimate TD-Error $\delta_t^{(i)}$ locally at node $i$ given by 
\begin{equation}
\delta_t^{(i)} = V(s_t^{(i)}; \mu) - r_t^{(i)} - \gamma V(s_{t+1}^{(i)}; \mu)
\end{equation}
The PPO agent has its weights trained with the nominal (anomaly-free) operations or simulations during offline mode. During training, each agent learns from locally-available information at time step $t$ due to the action taken at $t-1$, $a_{t-1}$. This information includes the number of acknowledgment packets (ACKs), $n_{t-1}$, $C/Q$-related route states (i.e. $\mathbf{c}_t$, $\mathbf{q}_t$ vectors, and their change rates $\Delta \mathbf{c}_t = \mathbf{c}_t - \mathbf{c}_{t-1}$, and $\Delta \mathbf{q}_t = \mathbf{q}_t - \mathbf{q}_{t-1}$) for the current node $i$. This data constitutes the input vector to the neural routing policy as 
\begin{equation}
\label{input_to_PPO_GAE_for_AD}
    s_t = \left[\mathbf{c}_t^{(i)}, \mathbf{q}_t^{(i)}, \Delta\mathbf{c}_t^{(i)}, \Delta\mathbf{q}_t^{(i)}, a_{t-1}^{(i)} \right].
\end{equation}
Since the reward is also used in the estimation of the TD-Errors, it follows that PPO handles locally-available rewards, amenable to real-time operations during execution.
\begin{equation}
\begin{split}
    r_t^{(i)} = & w_1 \cdot \mathsf{1}\!\left\{n_{t-1}^{(i)} = 1\right\} - w_2 \cdot \mathsf{1}\!\left\{n_{t-1}^{(i)} > 1\right\}(n_{t-1}^{(i)}-1) \\- & w_3 \cdot \mathsf{1}\!\left\{n^{(i)}_{t-1}= 0\right\} + w_4 \cdot  \mathsf{1}\!\left\{{D_{t-1}}^{(i)}\right\},
\end{split}
\end{equation}
where $\mathsf{1}\{x\}$ is equal to one if $x$ is true, and otherwise it is 0. $D_{t-1}^{(i)}$ is true if the current node $i$ just delivered a packet to its destination at the previous time step $t-1$ and has received an ACK from the destination for the packet. The weights $w_1, w_2, w_3, w_4$ are positive weights tune-able for the performance of the DeepCQ+ routing policy. During the execution, $\text{TD-Error}$ is computed from the current state $s_t^{(i)}$, next state $s_{t+1}^{(i)}$, and the reward $r_t^{(i)}$ which are available locally at the agent/node $i$. 

Of note, our routing policy training here is slightly different from the original DeepCQ+ \cite{kaviani2021deepcq+} as in this work we use locally available rewards. This means that we only reward the nodes that receive ACK for their delivery of packets and not all the nodes along the route. 

Next, we utilize the CUSUM-like method \cite{kurt2019sequential} for tracking and accumulating persistent outliers to detect anomalies in TD-Error data streams, as TD-Error is an individual metric with noisy sampling. Fig. \ref{fig:offline_online} shows our CUSUM-like anomaly detection with (a) the offline phase of training with nominal network data streams using PPO and (b) the online phase of executing the learned routing policy with CUSUM-like log-unlikelihood distribution and thresholds. Our CUSUM-like operation on TD-Error data streams is non-parametric, deriving the advantages of CUSUM with no need for knowing the probability density function of nominal TD-Error data and anomalous TD-Error data before and after the change point respectively. In fact, during training, TD-Error does not necessarily need to converge to zero. Nominal data streams can have non-zero TD-Error individually as a whole, as we need only to compute the deviation from possibly non-zero nominal TD-Error landscape to detect outliers. Our CUSUM-like operation over TD-Error data streams is suitable for real-time anomaly detection as it is sensitive to small shifts in sequential TD-Error data streams.

In the offline phase, we gather nominal TD-Error values $\tilde{\delta}_t^{(i)}$ from the anomaly-free scenario. Given this nominal dataset, we compute the nominal cumulative distribution function as an empirical probability distribution function (eCDF) of the nominal k-nearest-neighbor (kNN) summary statistics $\tilde{d}_t^{(i)}$ corresponding to the anomaly-free TD-Error $\tilde{\delta}_t^{(i)}$. From this eCDF $\tilde{f}(\cdot)$, we compute the corresponding tail probability ($p$-value), $p_t$. If a $\delta_t = \text{TD}(0)$ data point falls outside the acceptance region, that is, if its corresponding tail probability is less than $\alpha$ where $\alpha$ is user-defined acceptance level on the nominal empirical kNN distribution, we mark the TD-Error data point as an outlier. We continue to accumulate outliers to detect persistent ones as an anomaly. Specifically, if its tail probability $p$-value is less than $\alpha$, denoted as $p_{t}$, and formulated as ``log-unlikelihood'' $\ell_{t} = \log(\alpha/p_{t}$) (in lieu of the original CUSUM's log-likelihood ratio), an outlier is detected, and we use $\ell_{t}$ to update the decision function $g_{t}$.  Comparing the decision function $g_{t}$, which accumulates values $\ell_{t}$, and threshold $h$, the original CUSUM anomaly detection is formulated \cite{kurt2019sequential}, given $g_{0}$ = 0, as anomaly is called for $\Gamma = \text{Inf}_t \{g_{t} \geq  h\}$ and the decision function $g_t = \max(0,g_{t-1}+\ell_t)$. These offline and online operations endow DeepADMR with the capacity to see how effective DeepCQ+ routing policy works under environments outside trained regimes and in decentralized fashion.

\begin{figure}[t]
\centering
\includegraphics[width=0.485\textwidth]{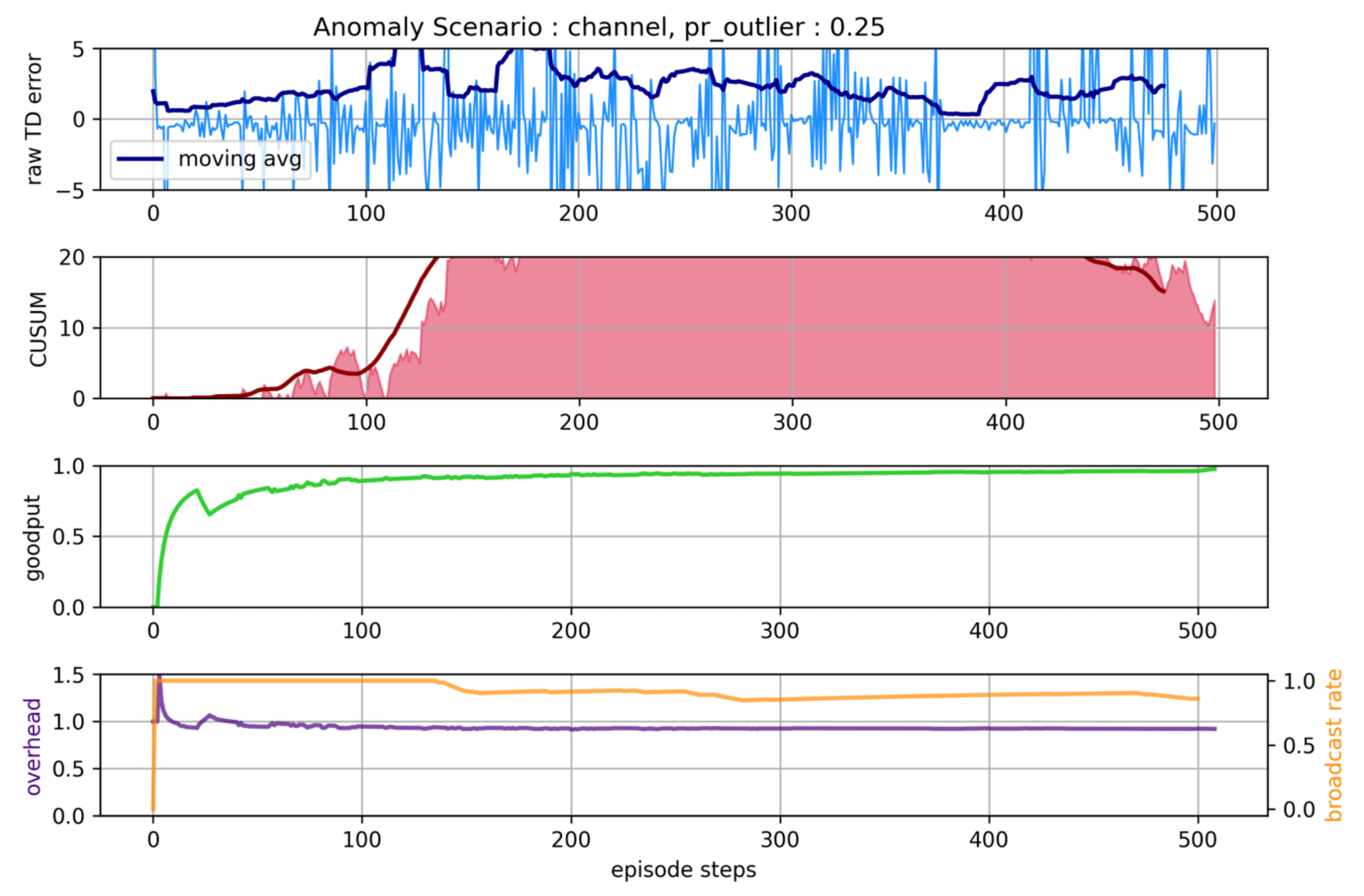}
\caption{\label{fig:TD_CUSUM_CH_AD} TD-Errors and TD-CUSUM output and network performance when channel disruption is present.}
\end{figure}

\section{DeepADMR Anomaly Detection Results}
We trained DeepCQ+ (DeepR2DN) for 12 nodes without any channel disruption or jammer in the RF environment. The average mobility speed (dynamic level), area size, and traffic flow rates are randomized within a specified range when DeepCQ+ routing policy is trained. This is considered our offline anomaly free operation. The DeepADMR is designed to detect severe and persistent deviation of our MANET environment and operation from the distribution of the scenarios that the DeepCQ+ is trained over. This does not include randomized dynamic levels, area size, and traffic flows but includes fundamental variations in the environment that cause unexpected behavior of the DeepCQ+ routing policy and therefore degraded performance. We tested DeepADMR for various network sizes and channel disruptions. 

\begin{figure}[t]
\centering
\includegraphics[width=0.485\textwidth]{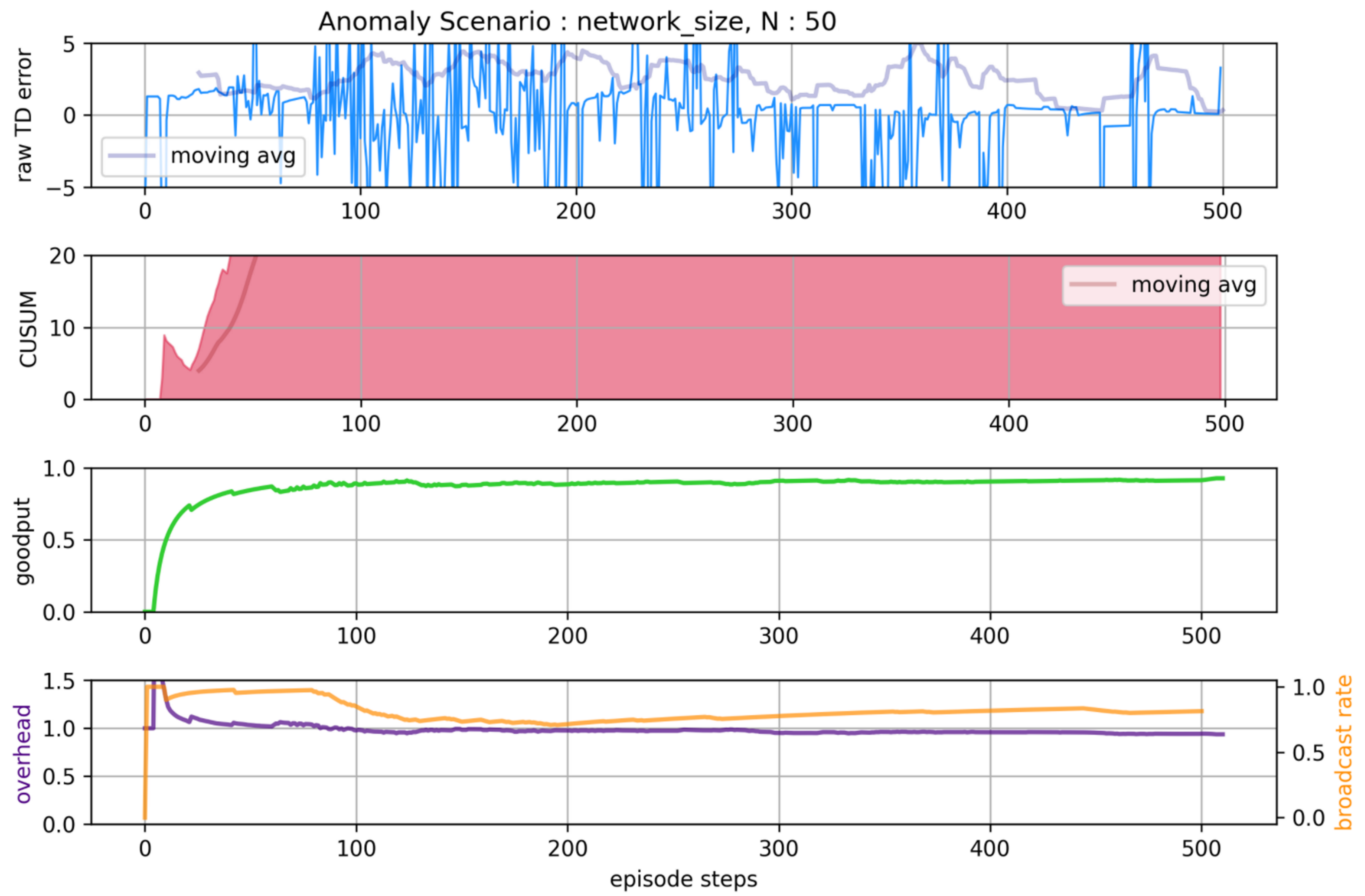}
\caption{\label{fig:TD_CUSUM_SIZE_50}TD-Errors and TD-CUSUM output and network performance for $N = 50$. The offline mode is $N = 12$.}
\end{figure}

\begin{figure}[ht]
    \centering
    \includegraphics[width=\linewidth]{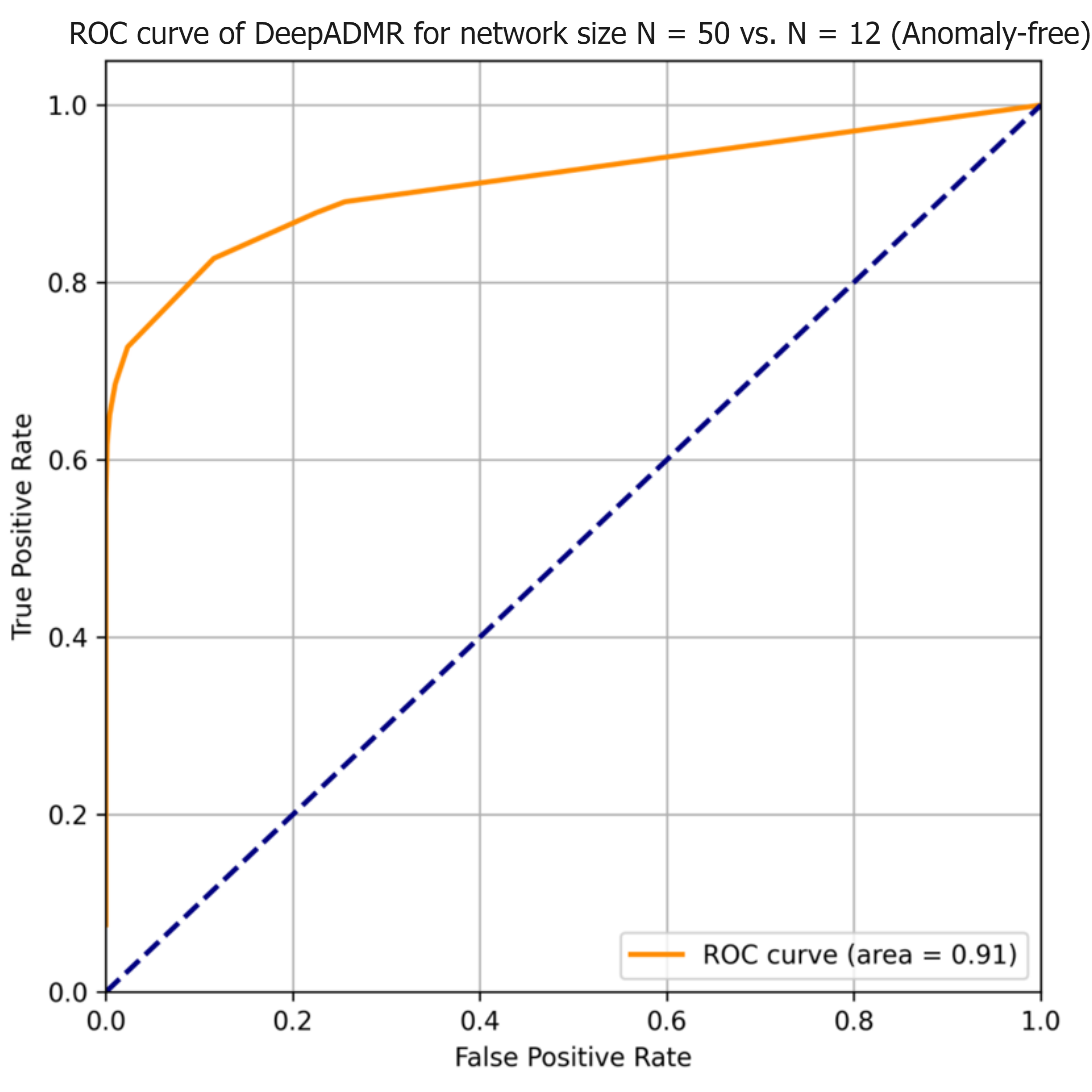}
    \caption{ROC curve for DeepADMR to detect anomaly scenario of network size 50 compared to the anomaly-free network size of 12 where the DeepCQ+ routing policy is trained over.}
    \label{fig:ROC_network_size}
\end{figure}

\begin{figure}[ht]
    \centering
    \includegraphics[width=0.485\textwidth]{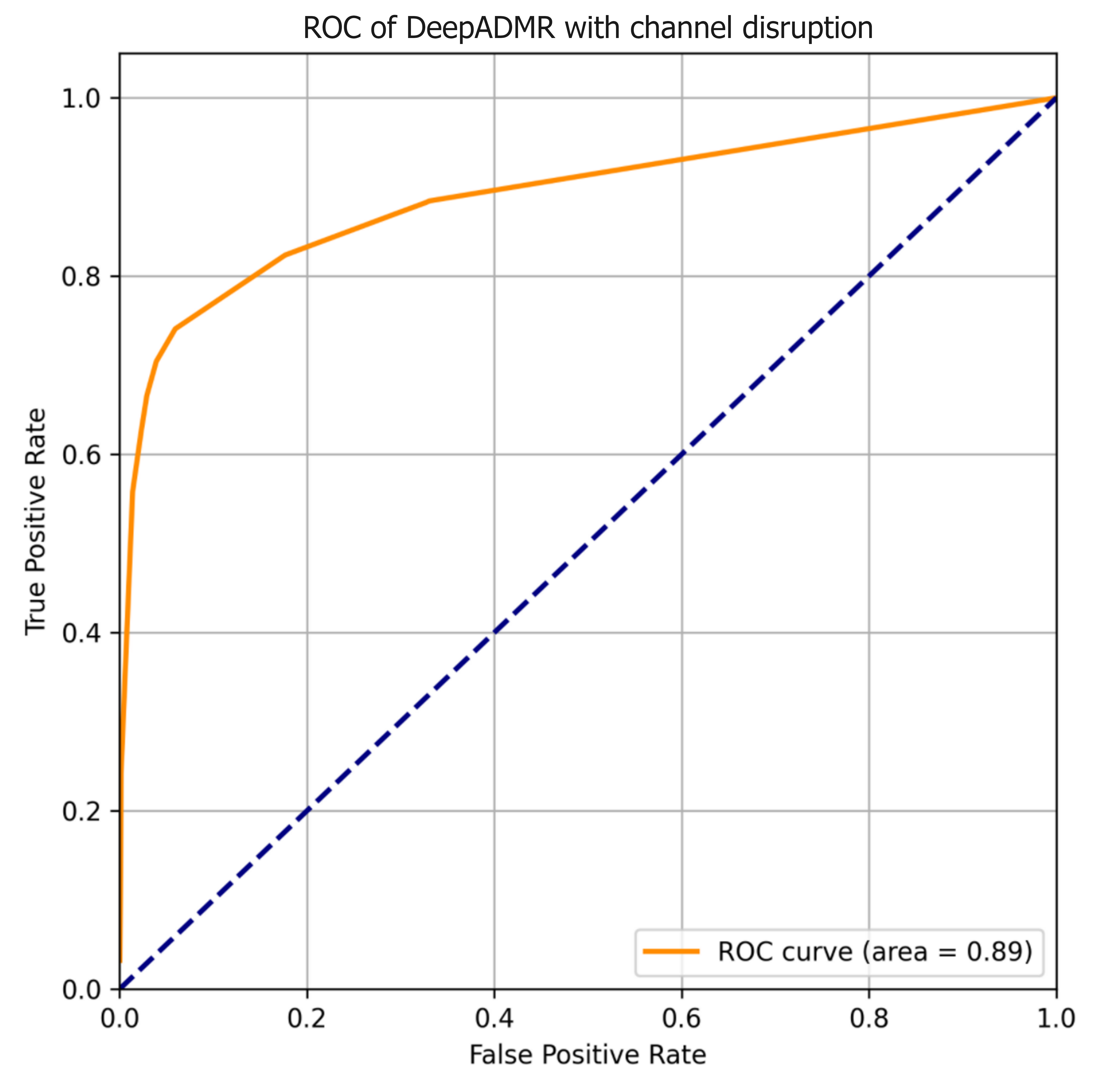}
    \caption{ROC curve for the DeepADMR to detect anomalies for the channel disruption or jammer (outlier RF environment)}
    \label{fig:ROC_CH}
\end{figure}

For anomaly detection, after training only for 12 nodes without channel disruption, we tested our DeepADMR neural anomaly detection. These anomaly scenarios include channel disruption present where the jammer is preset near the transmitting node but farther from receiving hop, preventing the transmitter to detect and decode the ACK packets. This causes unexpected observation at the transmitting node (e.g. no ACK is received). Another testing anomaly scenario is a much larger network size $N=50$ when trained for networks of size $N=12$. Although the performance of the DeepCQ+ shown to scale properly with network sizes, we still observe surprise outcome when execute it in very large network sizes. Note that the goal of the DeepADMR is to detect if the new scenarios strongly deviate from the training environment distributions. The results show that DeepADMR is effective to observe persistent TD-Error deviation (TD-CUSUM score $g_t$) computed by the CUSUM-like algorithm as shown in Fig. \ref{fig:TD_CUSUM_CH_AD} and Fig. \ref{fig:TD_CUSUM_SIZE_50}. The performance of the DeepADMR are measured in the form of receiver operating characteristics (ROC) curve and given in Fig. \ref{fig:ROC_network_size} and Fig. \ref{fig:ROC_CH}. 



\section{Conclusions}
We described DeepADMR, our neural anomaly detector based on CUSUM-like operations on TD-Error streams, as applied to the computer network routing anomaly detection problem. The results of DeepADMR anomaly detection on network routing problems caused by changing network size, various channel disruptions, changing mobility, and increased traffic indicated the effectiveness of our approach. Having reliable and trusted communications is indispensable in adversarial, dynamic, and uncertain environments, hence our DeepADMR manifests the core functionality of successful missions.

\section{Acknowledgements}
Research reported in this publication was supported in part
by Office of the Naval Research under the contract N00014-
19-C-1037. The content is alone the responsibility of the
authors and does not necessarily represent the official views of
the Office of Naval Research. The authors would like to thank
Dr. Santanu Das (ONR Program Manager) for his support and
encouragement.


\bibliographystyle{IEEEtran}
\bibliography{MARL_AD.bib}
\end{document}